# Non-perturbative renormalization of lattice QCD at all scales

Karl Jansen[a], Chuan Liu[a], Martin Lüscher[a], Hubert Simma[b], Stefan Sint[c], Rainer Sommer[d], Peter Weisz[c] and Ulli Wolff[e]

[a] Deutsches Elektronen-Synchrotron DESY,
   Notkestrasse 85, D-22603 Hamburg, Germany

[b] Department of Physics & Astronomy,
   The University of Edinburgh, Edinburgh EH9 3JZ, Scotland

[c] Max-Planck-Institut für Physik,
   Föhringer Ring 6, D-80805 München, Germany

[d] CERN, Theory Division, CH-1211 Genève 23, Switzerland

[e] Humboldt Universität, Institut für Physik,
   Invalidenstrasse 110, D-10099 Berlin, Germany

## Abstract

A general strategy to solve the non-perturbative renormalization problem in lattice QCD, using finite-size techniques and numerical simulations, is described. As an illustration we discuss the computation of the axial current normalization constant, the running coupling at zero quark masses and the scale evolution of the renormalized axial density. The non-perturbative calculation of $O(a)$ correction terms (as they appear in Symanzik's improvement programme) is another important field of application.

**1.** Renormalization in lattice QCD is intimately connected with the continuum limit and is thus of fundamental importance. In practical calculations, using numerical simulations, renormalization comes in when lattice data are related to quantities defined in the continuum theory. The additive and multiplicative renormalization of the quark masses in lattice QCD with Wilson quarks (chosen here) is an example of this, and there are very many more cases where renormalization constants need to be computed.

Perturbation theory is of limited value in this connection, because in general one is unable to compute more than two or three terms in the perturbation expansion. The numerical evaluation of the truncated series is hence liable to scheme ambiguities and the estimation of the truncation error becomes a matter of subjective judgement, particularly in situations where the coupling is large. Even if the series appears to converge well, some doubt will always remain that the result differs significantly from the exact number due to non-perturbative effects.

In some cases renormalization constants can be computed through numerical simulations. This has always been the method of choice for the additive quark mass renormalization. More recently various non-perturbatively defined renormalized gauge couplings have been calculated in the quenched approximation [1–8] and different ways to extract the renormalization constants associated with the isovector axial current and other composite fields have been described [9–15].

In this letter we propose to combine numerical simulations with finite-size techniques to solve the non-perturbative renormalization problem in lattice QCD. The method has previously been applied to compute the running coupling in SU(2) and SU(3) gauge theories [1–5]. Our aim here is to advertise its use in QCD by explaining the basic strategies in the context of this theory and by going through a list of applications.

**2.** When computing renormalization constants through numerical simulations, one is confronted with a number of technical difficulties. One of these arises from the fact that renormalization is scale dependent in general. It is then often necessary to trace the evolution of the renormalized parameters and renormalized composite fields from low to high energies, where contact can be made with perturbative renormalization schemes. For lattice gauge theory this presents a problem, since one cannot afford to simulate lattices covering physical scales orders of magnitude apart.

A second difficulty derives from the well-known limitation that simulations of QCD with very light quarks are usually prohibitively costly. As a result one is forced to adopt a mass dependent renormalization scheme or to rely on extrapolations to the chiral limit. Both options are not particularly attractive and it would be much preferable, if one were able to perform the parameter and field renormalizations directly at vanishing quark masses.

The influence of lattice effects is a further problem that should be addressed when discussing non-perturbative renormalization. A good example to illustrate this is the renormalization constant $Z_V$ of the (local) isospin current [9–14]. $Z_V$ may be fixed by requiring the isospin charge of some low-energy states to assume integer values. Depending on exactly which states are chosen to compute $Z_V$, results differing by terms of order $a$ (where $a$ denotes the lattice spacing) are obtained. Such ambiguities may be considered a systematic error on $Z_V$, but this point of view is not entirely satisfactory, since it is difficult to estimate the error in an objective manner. A better way to deal with the problem is to fix $Z_V$ through a definite normalization prescription and to study the approach to the continuum limit of the physical quantities involving the renormalized vector current that one is interested in.

Moving closer to the continuum limit means larger lattices and hence rapidly increasing cost. Less expensive ways to reduce cutoff effects include the use of an O($a$) improved action [16–20] along with improved composite fields [21]. A problem which one has here is that improvement needs to be verified. Moreover one may not be satisfied with a perturbative computation of the coefficients multiplying the O($a$) correction terms added to the action and the composite fields. For optimal improvement a non-perturbative determination of these coefficients may be necessary.

The finite-size technique discussed below is able to cope with all three problems mentioned in this section. This is rather non-trivial and we shall not attempt to present the method in full detail here. Instead we shall proceed from simple to more complex applications and defer all technical discussions to later publications.

**3.** We begin by considering QCD in a euclidean space-time volume of size $T \times L^3$, where the time-like extent $T$ and the spatial size $L$ are taken to be in a fixed proportion ($T = 2L$, for example). With suitable boundary conditions the renormalization of the theory in such a world proceeds as in ordinary space-time. In particular, the counterterms that must be added to the bare action may be chosen to be independent of $L$ and we shall assume that this is what has been done. From the point of view of the lattice theory, this simply means that the continuum limit at fixed $L$ (given in physical units) is obtained with the usual scaling of the bare coupling and quark masses.

A finite volume is usually regarded as a source of systematic error in numerical simulations of lattice QCD. Here, however, it is taken as a probe of the theory. To study renormalization and Symanzik improvement the relevant sizes $L$ are then smaller than 1 fm (hence the name "femto-universe" [22]). We shall in fact take $L$ to very small values when tracing the scale evolution of the renormalized parameters and fields to high energies.

Evidently the femto-universe is experimentally inaccessible and one may, therefore, be driven to conclude that it can only be of academic interest. This argumentation overlooks the fact that there are theoretical ways to relate finite volume with infinite volume physics, either through renormalized perturbation theory (at high energies) or via the bare parameters and fields on the lattice. It is then possible to arrange the computations so that all reference to a finite volume disappears from the final results. In other words, the femto-universe is only used as an intermediate device.

So far we did not specify the boundary conditions on the quark and gauge fields. Different boundary conditions amount to probing the theory in different ways, i.e. there are no fundamental reasons for choosing any particular prescription. To avoid inessential technical complications it is however wise to require that zero modes are excluded at tree-level of perturbation theory. A possible choice then are periodic boundary conditions in all spatial directions and Dirichlet boundary conditions at time $x_0 = 0$ and $x_0 = T$. Explicitly, at $x_0 = 0$ the gauge potential $A_\mu(x)$ is required to satisfy †

$$A_k(x) = C_k(\mathbf{x}), \qquad k = 1, 2, 3, \tag{3.1}$$

where $C_k$ is a given classical field. Since the Dirac equation is of first order, only half of the components of the quark and anti-quark fields $\psi(x)$ and $\overline{\psi}(x)$ can be prescribed, viz.

$$\tfrac{1}{2}(1 + \gamma_0)\psi(x) = \rho(\mathbf{x}), \qquad \overline{\psi}(x)(1 - \gamma_0)\tfrac{1}{2} = \overline{\rho}(\mathbf{x}). \tag{3.2}$$

Similar boundary conditions are imposed at $x_0 = T$ [23].

The euclidean functional integral, which may now be set up in the usual way, depends on the boundary values of the fields and may be interpreted as the quantum mechanical amplitude for going from the given field values at time $x_0 = 0$ to the values specified at time $x_0 = T$. It is, therefore, also referred to as the "Schrödinger functional". The renormalizability of the theory in these conditions has been studied in refs.[2,23,24] and it was found that no new counterterms are needed except for one term at the boundary which amounts to rescaling the boundary values $\rho$ and $\bar{\rho}$ by a logarithmically divergent renormalization constant. The renormalization of the coupling, the quark masses and the composite fields at times $0 < x_0 < T$ are not affected by the presence of the boundaries.

---

† For ease of presentation we use a continuum notation in this paper, but the formulae should be taken as shorthand for the corresponding lattice expressions

One of our motivations for choosing boundary conditions as specified above was to ensure that no zero modes occur to lowest order of perturbation theory. The eigenvalues of the free Dirac operator are in fact separated from zero by a gap which is proportional to the infrared cutoff $1/L$ at vanishing quark masses [23]. We have verified through numerical simulations, using a recently developed eigenvalue program [25], that the gap persists in the fully interacting theory for box sizes $L \leq 1$ fm. Moreover the fluctuations of the lowest eigenvalues remain small compared to the gap. The important consequence of this observation is that numerical simulations of the Schrödinger functional are feasible at zero quark masses. In the interesting range of $L$ such simulations are not significantly more expensive than simulations at small positive quark masses.

As an aside we remark that the Dirac operator in the continuum theory, in an arbitrary smooth background gauge field and with Schrödinger functional boundary conditions, can be shown to have no zero modes at vanishing quark masses. In particular, the Atiyah-Singer index theorem does not apply when such boundary conditions are imposed.

4. In the standard formulation of lattice QCD with two flavours of mass-degenerate Wilson quarks there are two bare parameters, the gauge coupling $g_0$ and the quark mass $m_0$. The renormalized quark mass $m_R$ is related to the bare mass through

$$m_R = Z_m(m_0 - m_c), \qquad (4.1)$$

where $Z_m$ and $m_c$ are the multiplicative and additive mass renormalization constants. We now describe how to compute $m_c$ using the femto-universe.

The idea is to consider the PCAC relation

$$\partial_\mu A_\mu^a(x) = 2m P^a(x) + \mathrm{O}(a) \qquad (4.2)$$

between the isovector axial current

$$A^a_\mu(x) = \overline{\psi}(x)\gamma_\mu\gamma_5 \tfrac{1}{2}\tau^a \psi(x) \tag{4.3}$$

and the associated (unrenormalized) density

$$P^a(x) = \overline{\psi}(x)\gamma_5 \tfrac{1}{2}\tau^a \psi(x) \tag{4.4}$$

($\tau^a$ denotes a Pauli matrix acting on the flavour indices of the quark fields). The mass $m$ appearing in eq.(4.2) is proportional to the renormalized quark mass. In particular, the critical bare mass $m_c$ may be calculated by finding the value of $m_0$ where $m$ vanishes.

For given bare parameters the mass $m$ can be extracted from ratios of correlation functions,

$$m = \tfrac{1}{2} \langle \partial_\mu A^a_\mu(x) \mathcal{O}^a \rangle \,/\, \langle P^a(x) \mathcal{O}^a \rangle + \mathrm{O}(a), \tag{4.5}$$

where $\mathcal{O}^a$ is a suitable polynomial in the quark and gluon fields supported in a region of space-time not containing $x$. Up to cutoff effects of order $a$, the computed value of $m$ should of course be independent of the source $\mathcal{O}^a$ and also of the volume and the boundary values of the fields ‡.

An interesting set of sources $\mathcal{O}^a$ may be constructed by differentiating the Schrödinger functional with respect to the boundary values of the quark field. This amounts to inserting quark fields at the boundary and so we introduce symbolic boundary fields $\zeta(\mathbf{x})$ and $\overline{\zeta}(\mathbf{x})$ through

$$\zeta(\mathbf{x}) = \frac{\delta}{\delta \overline{\rho}(\mathbf{x})}, \qquad \overline{\zeta}(\mathbf{x}) = -\frac{\delta}{\delta \rho(\mathbf{x})}. \tag{4.6}$$

In eq.(4.5) we now choose

$$\mathcal{O}^a = \int_0^L \mathrm{d}^3\mathbf{y}\, \mathrm{d}^3\mathbf{z}\, \overline{\zeta}(\mathbf{y})\gamma_5 \tfrac{1}{2}\tau^a \zeta(\mathbf{z}) \tag{4.7}$$

and set the boundary values $\rho$ and $\overline{\rho}$ to zero after differentiation. The boundary values of the gauge field are taken to be zero or constant abelian as in ref.[5].

---

‡ Recall that eq.(4.2) is a special case of the euclidean field equations. These are derived locally from the action and so do not depend on the boundary conditions (as long as one keeps away from the boundaries)

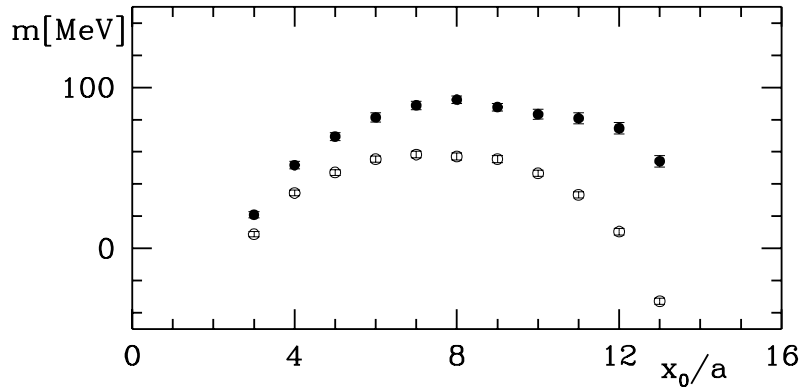

Fig. 1. Values of $m$ as computed from eq.(4.5), using the Wilson quark action. The open and full symbols correspond to zero and non-zero boundary values of the gauge field.

Note that the boundary fields in eq.(4.7) are projected to their zero momentum components. We do not need to choose a gauge to do this, since the gauge symmetry at the boundary is fixed by the boundary value of the gauge field.

The computation of the correlation functions in eq.(4.5) through numerical simulation is straightforward. A typical result from a $16 \times 8^3$ lattice with quenched Wilson quarks is shown in fig. 1. The bare mass $m_0$ is close to $m_c$ in this example and the bare coupling is such that $6/g_0^2 = 6.4$, which corresponds to a physical lattice size $L$ of about 0.4 fm [8,3]. Naively one would expect to see a plateau in this plot, since $m$ should be independent of the time $x_0$ at which the axial current is inserted. This is far from being the case. Moreover if we change the boundary values of the gauge field from zero to some weak field (half the strength of the field used in the computation of the running coupling in ref.[3]), the value of $m$ in the middle of the lattice is shifted by as much as 32 MeV.

Further studies reveal that these unexpected findings result from violations of chiral symmetry by lattice effects. The strongest effects (those of order $a$) can be canceled by using on-shell improved fields and an improved action. The latter is obtained by adding a Pauli term,

$$\frac{i}{4} c_{\text{sw}} a \sigma_{\mu\nu} F_{\mu\nu}(x), \qquad (4.8)$$

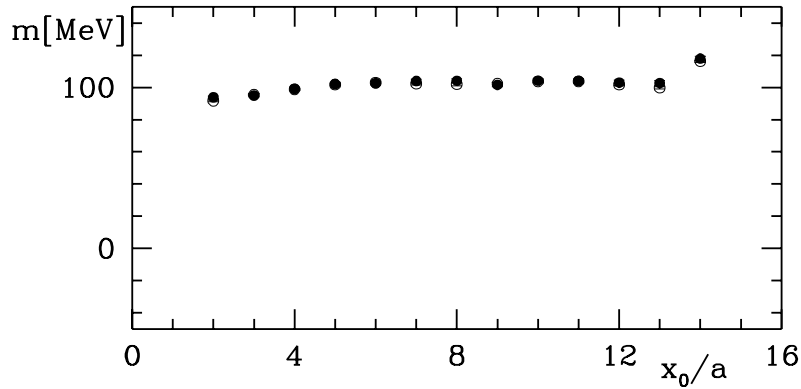

Fig. 2. Same as fig. 1, but using the O($a$) improved action and the improved axial current $\mathcal{A}^a_\mu/Z_A$ and density $\mathcal{P}^a/Z_P$ with $c_{\text{sw}} = 1.60$ and $c_A = -0.027$.

to the lattice Dirac operator, where $F_{\mu\nu}(x)$ denotes the gluon field strength [19,20]. For the renormalized on-shell O($a$) improved axial current and density we may take

$$\mathcal{A}^a_\mu = Z_A \left\{ (1 + b_A a m_q) A^a_\mu + c_A a \partial_\mu P^a \right\}, \qquad m_q = m_0 - m_c, \quad (4.9)$$

$$\mathcal{P}^a = Z_P (1 + b_P a m_q) P^a. \tag{4.10}$$

With these modifications PCAC is expected to be valid up to errors of order $a^2$, provided the improvement coefficients $c_{\text{sw}}, b_A, b_P$ and $c_A$ are chosen appropriately. To lowest order of perturbation theory $c_{\text{sw}} = b_A = b_P = 1$ and $c_A = 0$ [19,21].

The effect of improvement is quite dramatic (see fig. 2). There is now a wide plateau and the remaining dependence on the boundary values of the gauge field is barely significant. The O($a$) corrections associated with $b_A$ and $b_P$ are unimportant in the present context since they just amount to a rescaling of $m$ by a constant factor. They are anyway numerically insignificant at small quark masses and we have thus decided to neglect them in this calculation.

The quoted values of $c_{\text{sw}}$ and $c_A$ have been obtained by computing $m$ from three different correlation functions and adjusting the improvement coefficients until the same result is obtained in all three cases. In other words, the coefficients are determined by imposing a non-perturbative "improvement condition", which amounts to requiring PCAC to hold exactly in three correlation functions, with the same value of $m$. It should again be emphasized at this point that the improved action and fields are independent of the physical situation considered. Our results for the improvement coefficients are hence directly relevant for the computation of low-energy properties of QCD in large volumes.

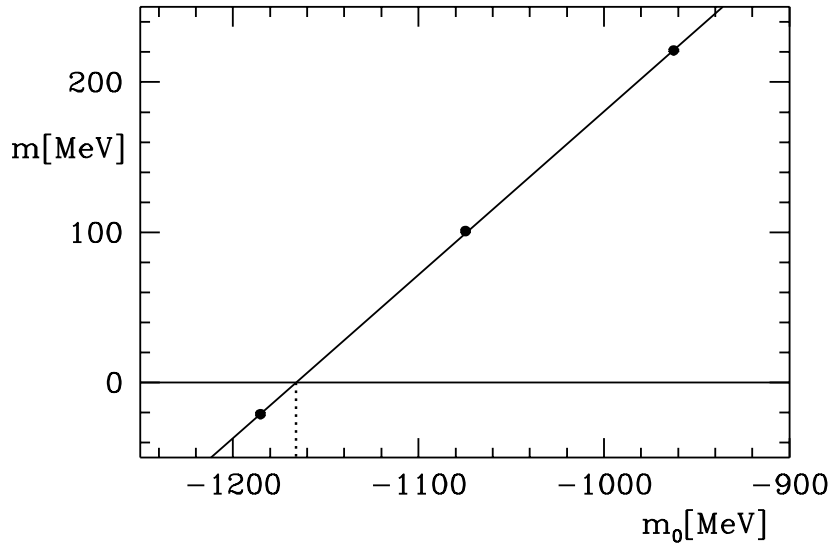

Fig. 3. Computation of the critical bare mass $m_c$ in the $O(a)$ improved theory. The lattice parameters are the same as in fig. 2. The three data points are interpolated linearly and $m_c$ is found at $m = 0$.

After improvement has been taken into account the calculation of the critical bare mass $m_c$ proceeds along the lines described earlier in this section. From the plateau in fig. 2 the mass $m$ is obtained with small errors. Repeating this calculation for two more values of the bare mass then yields the data points shown in fig. 3 and $m_c$ is finally obtained by linear interpolation. If there should be any doubt about the reliability of the interpolation, one can always verify that $m = 0$ within statistical errors at the computed value of $m_c$. The slope of the line in fig. 3 is about 1.09 and $m$ is hence close to $m_0 - m_c$ in the small quark mass region.

**5.** The axial current normalization constant $Z_A$ appearing in eq.(4.9) is determined by the chiral Ward identities. This has previously been discussed in a series of publications [9–13] and practical methods to extract $Z_A$ from the identities have been described, using numerical simulations.

The chiral Ward identities are derived locally from the QCD action and so are valid in finite volumes, too. The computation of $Z_A$ can thus be carried out in the femto-universe in essentially the same way as on physically large lattices. If we set the quark masses to zero (which simplifies the calculation considerably), the relevant Ward identity may be written in the form

$$\int_{\partial R} \mathrm{d}\sigma_\mu(x)\, \epsilon^{abc} \left\langle \mathcal{A}^a_\mu(x) \mathcal{A}^b_\nu(y) \mathcal{Q}^c \right\rangle = 2i \left\langle \mathcal{V}^c_\nu(y) \mathcal{Q}^c \right\rangle + \mathrm{O}(a^2). \qquad (5.1)$$

The integral in this formula is taken over the boundary of some space-time region $R$ containing the point $y$. $\mathcal{Q}^a$ is an arbitrary source located outside $R$ and $\mathcal{V}^a_\mu(x)$ denotes the renormalized on-shell improved isospin current,

$$\mathcal{V}^a_\mu(x) = Z_V \left\{ (1 + b_V a m_q) V^a_\mu + c_V a \partial_\nu T^a_{\mu\nu} \right\}, \qquad (5.2)$$

$$V^a_\mu(x) = \overline{\psi}(x)\gamma_\mu \tfrac{1}{2}\tau^a \psi(x), \qquad T^a_{\mu\nu}(x) = \overline{\psi}(x)\sigma_{\mu\nu} \tfrac{1}{2}\tau^a \psi(x). \qquad (5.3)$$

Note that the $\mathrm{O}(a)$ tensor correction to the vector current does not contribute to the isospin charge, but it could be important when computing $f_\rho$, for example.

A convenient choice for $R$ is the region between two equal time hyper-planes. For the source $\mathcal{Q}^a$ one may take

$$\mathcal{Q}^a = \epsilon^{abc} \mathcal{O}'^b \mathcal{O}^c, \qquad (5.4)$$

where $\mathcal{O}^a$ is given by eq.(4.7) and $\mathcal{O}'^a$ is constructed similarly at $x_0 = T$. If we set $\nu = 0$ in eq.(5.1), the correlation function on the right hand side reduces to a matrix element of the isospin charge between states with isospin 1 and so is determined by the normalization condition for the vector current. On the other side of the equation the normalization constant $Z_A$ is the only unknown coefficient, i.e. it can be computed by requiring eq.(5.1) to hold exactly for the chosen source and lattice parameters.

An attractive feature of this computation is that the fields in the Ward identity (5.1) are localized in different space-time regions. Cutoff effects from contact terms are hence avoided and the error can be argued to be of order $a^2$, if only the on-shell improved currents and the improved action are employed.

**6.** So far we have been exclusively concerned with scale independent renormalization constants. In this last section we address the more difficult problem of scale dependent renormalization. It is here that the proposed finite-size technique develops its full power.

Conventionally the renormalization of QCD is discussed in perturbation theory. In this framework the currently most popular renormalization scheme is the $\overline{\text{MS}}$ scheme of dimensional regularization. The associated renormalization scale $\mu$ is introduced in a rather implicit manner, but it is clear that it should be large compared to (say) the nucleon mass so that perturbation theory may be expected to apply.

At low energies, on the other hand, it is natural (and common practice in lattice QCD) to fix the bare parameters in the action by requiring that the nucleon mass and the masses of some pseudo-scalar mesons (one for each quark flavour) assume prescribed values. It is also possible to define renormalized composite fields by specifying some of their hadronic matrix elements. Such renormalization schemes are referred to as hadronic schemes.

An important part of the non-perturbative renormalization problem in lattice QCD is to determine the relation between the hadronic and the perturbative schemes. In particular, one would like to be able to compute the running coupling and the running quark masses at high energies given the nucleon and meson masses. One is also often confronted with the task of computing hadronic matrix elements of some local composite fields whose normalization is given at high energies through the $\overline{\text{MS}}$ scheme.

As already mentioned in sect. 2, the principal difficulty in such computations is to relate physical quantities defined at energy scales differing by factors of 10 to 100 (and may be more). In the following we describe in outline how this problem can be solved using finite-size techniques. The method has first been discussed in ref.[1] and was later applied in pure gauge theories [2–5].

The basic idea is to introduce an intermediate finite-volume renormalization scheme, where all renormalized quantities are defined at scale $1/L$. The Schrödinger functional provides a technically attractive framework to set up such a renormalization scheme, which is then called the SF scheme. Using perturbation theory the SF scheme can be matched with the $\overline{\text{MS}}$ scheme when $L$ is sufficiently small and $\mu L$ of order 1. At the lower end of the energy scale, the SF scheme can be related to the hadronic schemes through numerical simulations. The crucial point now is that the scale evolution in the SF scheme is computable through a non-perturbative recursive procedure. Via the SF scheme one is hence able to connect the $\overline{\text{MS}}$ scheme with the hadronic schemes.

We now need to be a bit more explicit on the definition of the SF scheme and shall then explain how the scale evolution of the renormalized parameters and fields can be computed. We choose to set up the SF scheme in such a way that all normalization conditions are given at zero quark mass [26].

The renormalized coupling $\alpha_{\rm SF}(q)$, $q = 1/L$, is obtained from the Schrödinger functional by calculating the response of the system to changes of the boundary values of the gauge field (full details are given in refs.[3,24]). We set $m_0 = m_c$ in this definition, as indicated above, and scale the chosen boundary values proportionally to $L$. The only variables on which $\alpha_{\rm SF}(q)$ depends are then the bare coupling $g_0$ and the lattice size $L/a$. It has been shown that $\alpha_{\rm SF}(q)$ is a renormalized quantity which satisfies the expected massless renormalization group equation. Moreover its relation to $\alpha_{\overline{\rm MS}}(q)$ has been worked out to one-loop order [24].

For the running quark mass we may take

$$m_{\rm SF}(q) = mZ_{\rm A}/Z_{\rm P}, \qquad q = 1/L, \tag{6.1}$$

where $m$ is determined from PCAC using the (unrenormalized) improved current $\mathcal{A}^a_\mu/Z_{\rm A}$ and density $\mathcal{P}^a/Z_{\rm P}$. The computation of the axial current normalization constant $Z_{\rm A}$ has been discussed in the previous section. To fix the renormalization constant $Z_{\rm P}$ of the axial density (4.10), we must impose a normalization condition. A simple possibility is to require that

$$\langle \mathcal{P}^a(x)\mathcal{O}^a \rangle^2 = -(9/L^6)\langle \mathcal{O}'^a \mathcal{O}^a \rangle, \tag{6.2}$$

at zero quark mass, zero boundary values, $x_0 = T/2$ and $T = 2L$. The proportionality factor in this equation has been chosen so that $Z_{\rm P} = 1$ to lowest order of perturbation theory. Other composite fields can be renormalized in a similar manner.

The evolution of $\alpha_{\rm SF}(q)$ from some large initial value of $q$ towards lower scales may now be calculated simply by increasing $L$ at fixed bare parameters. Of course we can only increase $L$ by a factor of 2 or so, as otherwise one ends up with lattices that cannot be simulated with the currently available computers. In a second step the lattice spacing $a$ is increased at fixed renormalized parameters. This amounts to lowering $L/a$ by say a factor of 2 and adjusting the bare parameters so that the quark mass (equal to zero) and the renormalized coupling remain unchanged. Now we can evolve the coupling by another factor of 2, and so on, until the low-energy regime is reached.

With little additional effort one can also compute the scale evolution of the renormalization constant $Z_{\rm P}$ (which in turn determines the evolution of the running quark mass). In the recursion we simply have to calculate the ratios $Z_{\rm P}(g_0, L'/a)/Z_{\rm P}(g_0, L/a)$ when the box size is increased from $L$ to $L'$ at fixed bare parameters. The product of all these ratios is equal to the total change in the normalization of the renormalized axial density during the evolution. It should be emphasized that the lattice cutoff $1/a$ is always much greater than the renormalization scale, at all stages of the calculation. Cutoff effects are hence expected to be small. Moreover, following refs.[1–5], any remaining effects can be extrapolated away by simulating sequences of lattices with decreasing lattice spacings.

Before the running coupling and quark mass can be computed along the lines explained above, some preparatory work is still required (simulation algorithms for dynamical quarks with $O(a)$ improved action need to be developed, for example). As an intermediate step one may be interested to perform simulations with zero and negative numbers of dynamical quark flavours [27,28], since these are relatively cheap and thus provide a useful laboratory to study some of the technical issues involved. Such calculations may also give interesting additional insights into the dependence of the evolution of the renormalized parameters and fields on the number of light quarks.

We are indebted to Roberto Frezzotti and Chris Sachrajda for clarifying discussions on the improved axial current. The numerical simulations have been performed on the powerful APE/Quadrics computers at DESY-IfH (Zeuthen). We thank the staff of the computer center at Zeuthen for their support.